\documentclass[aps,pra,floatfix,twocolumn,superscriptaddress,showpacs]{revtex4-1}

\usepackage{graphicx}
\usepackage{braket}
\usepackage{amsmath}
\usepackage{amssymb}

\newcommand{\comm}[2]{\left[#1,#2\right]}

\begin{document}

\author{J.~Schachenmayer}
\affiliation{JILA, NIST, Department of Physics, University of Colorado, 440 UCB, Boulder, CO 80309, USA}
\author{L.~Pollet}
\affiliation{Department of Physics, Arnold Sommerfeld Center for Theoretical Physics and Center for NanoScience, University of Munich, 80333 Munich, Germany}
\author{M.~Troyer}
\affiliation{{Theoretische Physik, ETH Zurich, 8093 Zurich, Switzerland}	}
\author{A.~J.~Daley}
\affiliation{Department of Physics and SUPA, University of Strathclyde, Glasgow G4 0NG, UK}
\affiliation{Department of Physics and Astronomy, Department of Physics, University of Pittsburgh, Pittsburgh, PA 15260, USA}

\title{Thermalization of strongly interacting bosons after spontaneous emissions in\\ optical lattices}

\begin{abstract}
We study the out-of-equilibrium dynamics of bosonic atoms in a 1D optical lattice, after the ground-state is excited by a single spontaneous emission event, i.e.~after an absorption and re-emission of a lattice photon. This is an important fundamental source of decoherence for current experiments, and understanding the resulting dynamics and changes in the many-body state is important for controlling heating in quantum simulators. Previously it was found that in the superfluid regime, simple observables relax to values that can be described by a thermal distribution on experimental time-scales, and that this breaks down for strong interactions (in the Mott insulator regime). Here we expand on this result, investigating the relaxation of the momentum distribution as a function of time, and discussing the relationship to eigenstate thermalization. For the strongly interacting limit, we provide an analytical analysis for the behavior of the system, based on an effective low-energy Hamiltonian in which the dynamics can be understood based on correlated doublon-holon pairs.
\end{abstract}

\maketitle


\section{Introduction}
\label{intro}
An important fundamental question in many-body quantum mechanics is to
what extent and under which conditions an isolated system perturbed away from equilibrium will undergo
thermalization, in the sense that
at long times the system will reach a steady state where simple
observables equal the values for a thermal distribution
\cite{deutsch_quantum_1991,Srednicki1994,Rigol2008,Cazalilla2011,Rigol2012}. Recently it has been
possible to observe integrable dynamics for strongly interacting cold
gases confined to move in one dimension \cite{Kinoshita2006}, and thus reach regimes where systems either do not thermalize in this standard sense \cite{Rigol2009} or otherwise undergo
generalized thermalization \cite{Cassidy2011,Rigol2011}. 

These fundamental questions also have an important impact on the application of cold atoms as quantum simulators \cite{Bloch2012,Cirac2012}. Development in experiments with these systems has reached a stage where it is possible to tailor interesting many-body Hamiltonians, and study the many-body physics of corresponding ground states. However, some of the most interesting physical regimes require the realization of states with small energy gaps, requiring exceptionally low temperatures and entropies, which provide a key challenge for current experiments \cite{McKay2011}. This has been particularly true in attempts to observe quantum magnetism in optical lattices within strongly interacting regimes $|U|\gg J$, where $U$ is the on-site interaction energy, and $J/\hbar$ indicates the tunneling amplitude for particles moving between neighboring sites. There, magnetic order in multi-species mixtures is driven by terms of the order of $J^2/U$, which is typically small for current experiments  \cite{Jordens2010,McKay2011,Hofstetter2002,Mathy2012,Jordens2008,Paiva2010,De-Leo2011,Fuchs2011}. 

In this context, it is very important to understand and control the many-body effects of competing heating processes in experiments. Previously, it has often been assumed that all of the energy added to the
system will be thermalized, causing an effective increase in temperature. However, in regimes in which the system does not thermalize excitations entirely (or for some types of excitations, at all), the behavior is more complex. A simple example where excitations cannot thermalize on typical experimental timescales is given by  spontaneous emission events (incoherent light scattering). In such processes, the main contribution to the increase in the average energy of the system as a function of time comes from atoms being excited to higher bands of an optical lattice \cite{Gerbier2010,Pichler2010,Gordon1980,Dalibard1985}. Because the bandgap is usually much larger than the typical energy scales $U$ and $J$ of dynamics in the lowest band, it is not possible for the system to thermalize most of the energy added to the system as a function of time. 

At the same time, understanding this process is made more complex by the fact that processes exciting particles to higher bands are rare compared with processes leaving particles in the lowest band in typical experiments \cite{Gerbier2010,Pichler2010}. As a result, the dynamics can actually be dominated by heating in the lowest band of the lattice, the basic effects of which have been studied in several recent articles for bosons \cite{Gerbier2010,Pichler2010,schachenmayer_spontaneous_2014,Poletti2012,barmettler_propagation_2012,Poletti2013,sorg_relaxation_2014} and for fermions \cite{Bernier2013,Sarkar2014,Bernier2014}. It is then natural to ask, in particular, whether the system can thermalize after spontaneous emission events that leave atoms in the lowest band.  

In this article, we study the interplay between spontaneous emissions and thermalization in detail, asking under what conditions key simple quantities such as the quasimomentum distribution and the kinetic energy relax to values given by a thermal distribution. In each case, we ask this question in a practical context by fixing a thermal distribution such that we take a canonical ensemble with the temperature $T$ chosen so that the expectation value of the energy in the thermal distribution matches the expectation value of the energy after a spontaneous emission event. In the case of strong interactions, we might expect that the system reaches integrable limits where the system does not thermalize. On the other hand, one might also expect that as in general, the Bose-Hubbard Hamiltonian is non-integrable, that outside the special cases of strong interactions ($U/J\rightarrow \infty$) and non-interacting systems ($U=0$), thermalization would essentially always occur. 

Instead, we find that this generalization is not universally applicable. It is important to note that spontaneous emissions correspond to local quenches for the many-body state, leading to population of low-lying excited states. As we showed previously \cite{schachenmayer_spontaneous_2014}, after a spontaneous emission, depending 
on the parameter regime and the corresponding different characteristics of the low-energy spectrum: either (i) the system relaxes over short times to 
thermal values of the quasimomentum distribution and kinetic energy, or (ii) on short timescales, the system relaxes to states
that are clearly non-thermal, even if all atoms remain in the lowest Bloch band. The latter regime was found to occur if the interactions are strong enough so that the ground-state of the system corresponds to a Mott insulator state. Below we elaborate on these results especially for the quasimomentum distribution, and discuss their interpretation in terms of eigenstate thermalization.

In the limit where the system is close to an ideal Mott-insulator (a product state with a fixed number of particles per site) and in the case of unit filing, one can describe the dynamics of the system in an effective model, where one restricts the local Hilbert space to having a maximum of two particles per site. The low-lying excitations can then be described as correlated pairs of holes (``holons'') and doubly occupied sites (``doublons'')  \cite{barmettler_propagation_2012,cheneau_light-cone-like_2012}. Here we use this approximation to show that no thermalization is present for spontaneous emission in this limit. We compare our result to exact numerical calculations either using exact diagonalization in small systems, or by making use of t-DMRG techniques \cite{Vidal2004,Daley2004,White2004,Verstraete2008}.

This article is organized as follows. We begin by introducing spontaneous emissions and thermalization in these systems in Section~\ref{sec:spont_em_therm}. Then, in Section~\ref{sec:eth} we review the explanation of the basic behavior we find in terms of the eigenstate thermalization. In Section~\ref{sec:dh} we discuss how the behavior can be understood in the strongly interacting limit, making predictions that could be observed for the propagation of excitations when the system does not thermalize in that limit. In Section~\ref{sec:summary} we then provide a summary and outlook.

\section{Thermalization of spontaneous emissions}
\label{sec:spont_em_therm}

Spontaneous emissions are a fundamental source of heating in optical
dipole potentials \cite{Gordon1980,Dalibard1985}, and one of the key
contributing heating sources in current experiments with cold atoms in
optical lattices \cite{Gerbier2010,Pichler2010}. As shown in Ref.~\cite{Pichler2010}, a multi-band many-body master equation can be derived to
describe this process in the case of far detuned lattice light. An adiabatic elimination of the electronic excited states leads to an equation for atoms in their electronic ground states. Taking this, and assuming the typical experimental case in which the lattice spacing $a$ is comparable to the wavelength of
scattered photons, $a = \lambda/2$, the dynamics of the
many-body density operator $\rho$ follows a master equation of the form ($\hbar\equiv 1$), $\dot\rho=-{\rm i}[H_{\rm MB},\rho]+\mathcal{L}\rho$. Here $H_{\rm MB}$ is a multi-band Bose-Hubbard Hamiltonian and $\mathcal{L}\rho$ includes dissipative Lindblad terms with on-site jump operators $\propto (b_i^{[n]})^\dag  b_i^{[m]}$. Here, $b_i^{[n]}$ is the bosonic annihilation operator at site $i$, for a particle in band $m$.

For relatively deep optical lattices, transition rates for inter-band processes coupling neighboring Bloch bands are strongly
suppressed by the square of the Lamb-Dicke parameter, $\eta=2\pi
a_T/\lambda$.  For a typical experiment with a lattice depth
$V_0\approx 8E_R$ [$E_R=4\pi^2 \hbar^2/(2 m \lambda^2)$, where $m$ is
  the mass of the atom], $\eta^2\sim 0.1$. In the usual case of a red-detuned optical lattice
  the dominant processes are thus intra-band processes, which return the atoms to their initial Bloch band. Provided we initially consider the system to be in the lowest band, processes
  accessing higher bands are suppressed by a factor of the order $\eta^4$ or greater, and the master equation for the 1D model simplifies to:
\begin{align}
\dot\rho=-i\comm{H}{\rho}-\frac{\gamma}{2}\sum_{{i}}
\comm{n_{{i}}}
{\comm{n_{{i}}}{\rho}}, \label{eq:meq}
\end{align}
where  $n_i=b_i^\dag b_i$ is the particle number operator for the lowest band, and $\gamma$ is the scattering rate. The Hamiltonian is the usual single band Bose-Hubbard Hamiltonian
\begin{align}
H=-J\sum_{i} \left( b_{{i}}^\dag b_{i+1} +  b_{{i+1}}^\dag b_{i} \right)+ \frac{U}{2} \sum_i n_i(n_i-1).
\label{eq:bh}
\end{align}
In the following we will focus on a system with $M$ sites and $N=M$ particles, so that the filling is $\bar n=N/M=1$. In this case the system undergoes a ground-state quantum phase transition between a superfluid (SF) and a Mott insulating (MI) state at the critical value, which was estimated to be $U_c \approx 3.25(5) J$ \cite{Schonmeier-Kromer2014,kuhner_one-dimensional_2000}.
The underlying physics of the lattice-photon scattering process can  be summarized as the environment obtaining information about the position of an atom. Thereby the atoms are localized on the length-scale of a single lattice-site size. 

\begin{figure}[tb]
 \includegraphics[width=1\columnwidth]{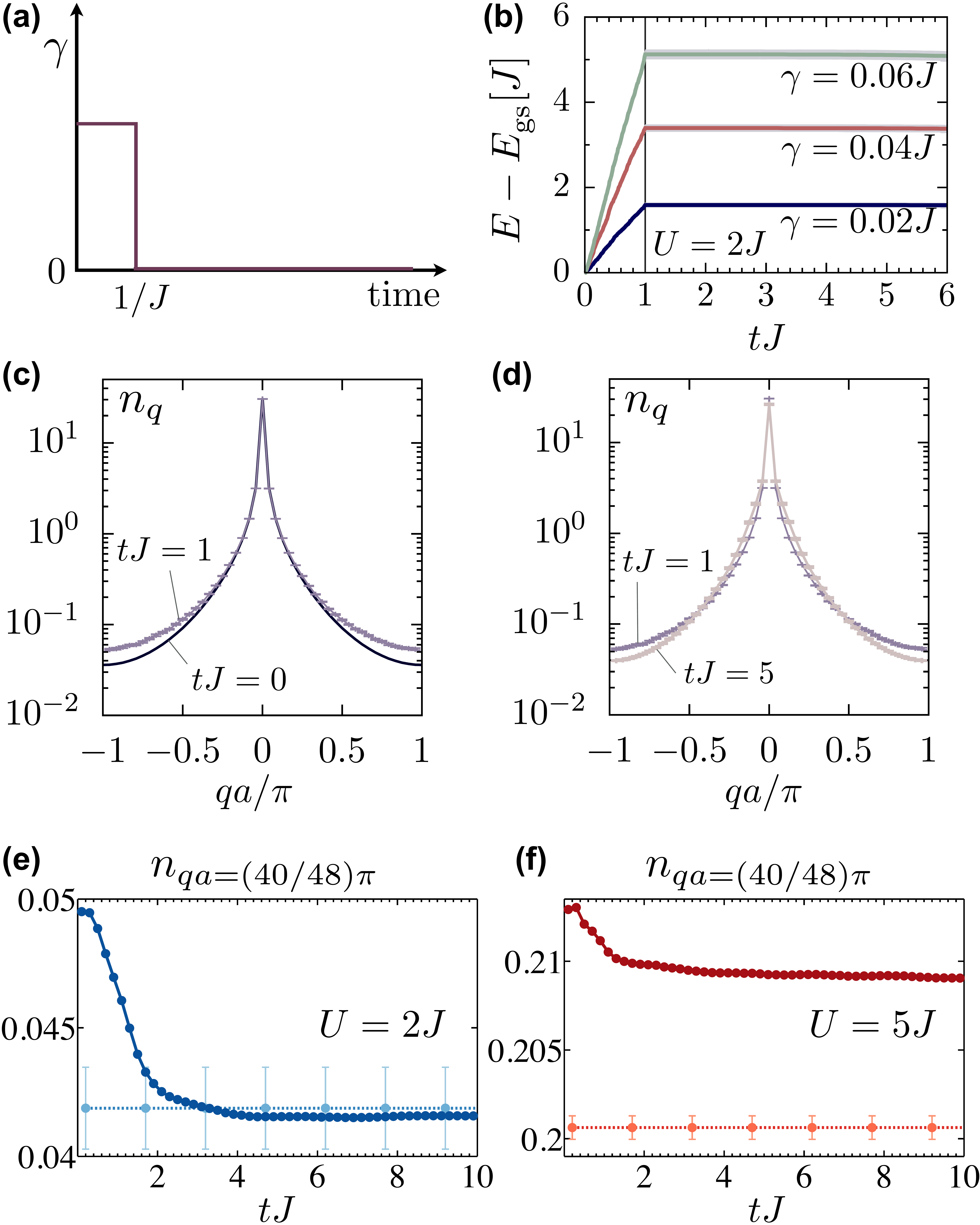}
\caption{{\it Example of heating and relaxation dynamics--}(a) A system with $N=48$ particles on $M=48$ sites and $U=2J$ is initially in its  ground state and subjected to heating with different scattering rates $\gamma$ for a short time. (b)  Depending on $\gamma$ this leads to an increase in energy. (c) Evolution of the quasi-momentum distribution $n_q$ during the heating period ($\gamma=0.06J$). The tails are lifted due to the localization of particles. (d) Relaxation dynamics of $n_q$ after the heating is switched off. The system relaxes to a broadened distribution at $tJ=5$ (t-DMRG results, bond-dimension, $D=256$, $500$ quantum trajectories). (e/f) Relaxation of a large $q$ component of the quasi-momentum distribution after a single spontaneous emission on an arbitrary site (we consider the a weighted ensemble average of jumps on the different sites). The dashed horizontal line indicates the equilibrium value (Monte-Carlo calculation). Thermalization only occurs in the superfluid regime (e), and not in the Mott insulating regime (f) [t-DMRG results, bond-dimension, $D=512,256$ for (e),(f)].}
\label{fig:mdf_examples} 
\end{figure}

Numerically we can fully simulate the master equation \eqref{eq:meq} also for large 1D systems by combining t-DMRG techniques with a Monte-Carlo sampling over quantum trajectories \cite{Daley2014,Molmer1993,Gardiner2005,Carmichael1993}. There, the density matrix is obtained from a statistical average over many trajectories with randomly applied jump operators \cite{Daley2014}. To give an example, in Fig.~\ref{fig:mdf_examples} we plot the dynamics for the evolution under equation~\eqref{eq:meq}. We start in the ground-state for a system with $U=2J$ (SF regime, $M=48$). Then we turn on light scattering for a short time up to $tJ=1$ and for different values of $\gamma$ (illustrated in Fig.~\ref{fig:mdf_examples}a). This leads to an increase in energy (Fig.~\ref{fig:mdf_examples}b). Once the heating is switched off, the system then relaxes. This relaxation is for example seen when looking at the evolution of the quasi-momentum distribution
\begin{align}
 n_q=\frac{1}{M}\sum_{n,m} e^{-iq(n-m)} \langle b^\dag_n b_m \rangle.
\end{align}
During the heating period the localization of particles leads to a ``lifting'' of tails of this distribution, as seen in Fig.~\ref{fig:mdf_examples}c. Once the heating is switched off, the high quasi-momentum components become redistributed and the system is relaxing to a broadened distribution, depicted in Fig.~\ref{fig:mdf_examples}d. The question whether the system thermalizes or not depends now on the shape of this distribution. In the case that the steady state distribution coincides with a Boltzmann distribution for a temperature corresponding to the same mean energy, we call the system thermalized, otherwise we don't. Obviously, the question of thermalization is connected to the observable under consideration. In Ref.~\cite{schachenmayer_spontaneous_2014} it was found that for example the tails of the quasi-momentum and the kinetic energy $E_{\rm kin} = \langle  H_{\rm kin} \rangle= - 2 J \sum_i \langle  b_{{i}}^\dag b_{i+1} \rangle+ \langle  b_{{i+1}}^\dag b_{i}  \rangle$ thermalize on experimentally relevant time-scales in the SF regime and that this thermalization breaks down in the MI. 

Here, we demonstrate a specific example of this in Figs.~\ref{fig:mdf_examples}e,f. There we show relaxation of a high-$q$ component of the quasimomentum distribution, $q=40\pi/48$, in a system with $M=48$ lattice sites. We can clearly see that after a short period of time, this component relaxes to its value in a corresponding thermal distribution provided that we initially start in a superfluid state with $U/J=2$. In the MI limit, we see that the behavior is similar, in so far as there is a rapid relaxation of this value of quasimomentum on short times. However, this leads to a value that is markedly different to the corresponding thermal value. It is naturally possible that in this regime, there are processes on very long timescales that will eventually lead to thermalization of this quantity. But it is clear that on the MI side of the phase transition, the system exhibits a qualitatively different behavior in terms of how close it is to thermal distributions on the scale of a few tunnelling times. This will significantly impact the ability of the system to thermalize energy added by heating on the timescales of the heating process itself. 

It is important to note that in the regime where the system is relaxing, the thermalization and all of the corresponding dynamics is determined by unitary closed system dynamics alone. In the remainder of this paper we will simplify the question of thermalization to the coherent dynamics after a single spontaneous emission event, i.e.~we ask whether expectation values of observables relax to thermal values in a unitary time-dynamics after we apply a single quantum jump. Thus, we consider a unitary time-evolution of an initial state
\begin{align}
\ket{\psi_{l}(t=0^+ )}=\frac{ n_{l}\ket{\psi_g}}{|| n_{l}\ket{\psi_g}||},
\end{align}
where $\ket{\psi_g}$ denotes the ground-state of the system, and then take a stochastic average over the sites $l$, which is weighted proportional to the square of the density on site $l$, $\bra{\psi_g} n_l^2 \ket{\psi_g}$.
The mechanism of this thermalization of closed quantum systems can be explained via eigenstate thermalizaton, a mechanism which we will review in the next section.

\section{Eigenstate thermalization}
\label{sec:eth}

The question whether unitary time-dynamics leads to thermalization or not can be re-phrased in terms of the ``eigenstate thermalization''  \cite{deutsch_quantum_1991,Srednicki1994,Rigol2008}, which we will briefly review here. Consider the situation where the initial state is not an eigenstate of the Hamiltonian (such as it is the case after a spontaneous emission). The state can be expanded into energy eigenstates defined via $H\ket{\alpha}=E_\alpha\ket{\alpha}$, $|\psi\rangle = \sum_\alpha c_\alpha | \alpha \rangle$, and the time-evolution is thus given by
\begin{align}
	|\psi(t)\rangle = \sum_\alpha {\rm e}^{-\frac{i}{\hbar} E_{\alpha} t } c_\alpha | \alpha \rangle 
	.
\end{align}
Then the time-dependent expectation value of any observable $\hat O$ is
\begin{align}
\langle \hat O \rangle_t = \langle \psi(t) |\hat O |\psi(t) \rangle = \sum_{\alpha,\beta}  {\rm e}^{-\frac{i}{\hbar} (E_{\alpha} - E_\beta) t } c_\beta^* c_\alpha \langle \beta | \hat O | \alpha \rangle.
\end{align}
If the system relaxes into a steady state, this steady state must be identical to the long-time average
\begin{align}
\overline{\langle \hat O \rangle_t } = \lim_{T \to \infty} \frac{1}{T} \int_0^T dt \langle \psi(t) |\hat O |\psi(t) = \sum_{\alpha} |c_\alpha|^2 \langle \alpha | \hat O | \alpha \rangle .
\end{align}
Thus, the value of the steady state expectation (if such a steady state is developed) only depends on two variables: i) The expansion coefficients of the initial state, $c_\alpha$, and ii) the expectation values of the energy-eigenstates with the particular observable, $\bra{\alpha} \hat O \ket{\alpha}$. The eigenstate thermalization hypothesis states, that if the the eigenenergy expectation values of an observable vary smoothly in the energy window defined by the $c_\alpha$ (and the off-diagonal $\bra{\alpha} \hat O \ket{\beta}$, $\alpha\neq \beta$ are small), the system relaxes into a state for which the observable can be described with a micro-canonical ensemble  \cite{deutsch_quantum_1991,Srednicki1994}.

\begin{figure}[t]
 \includegraphics[width=1\columnwidth]{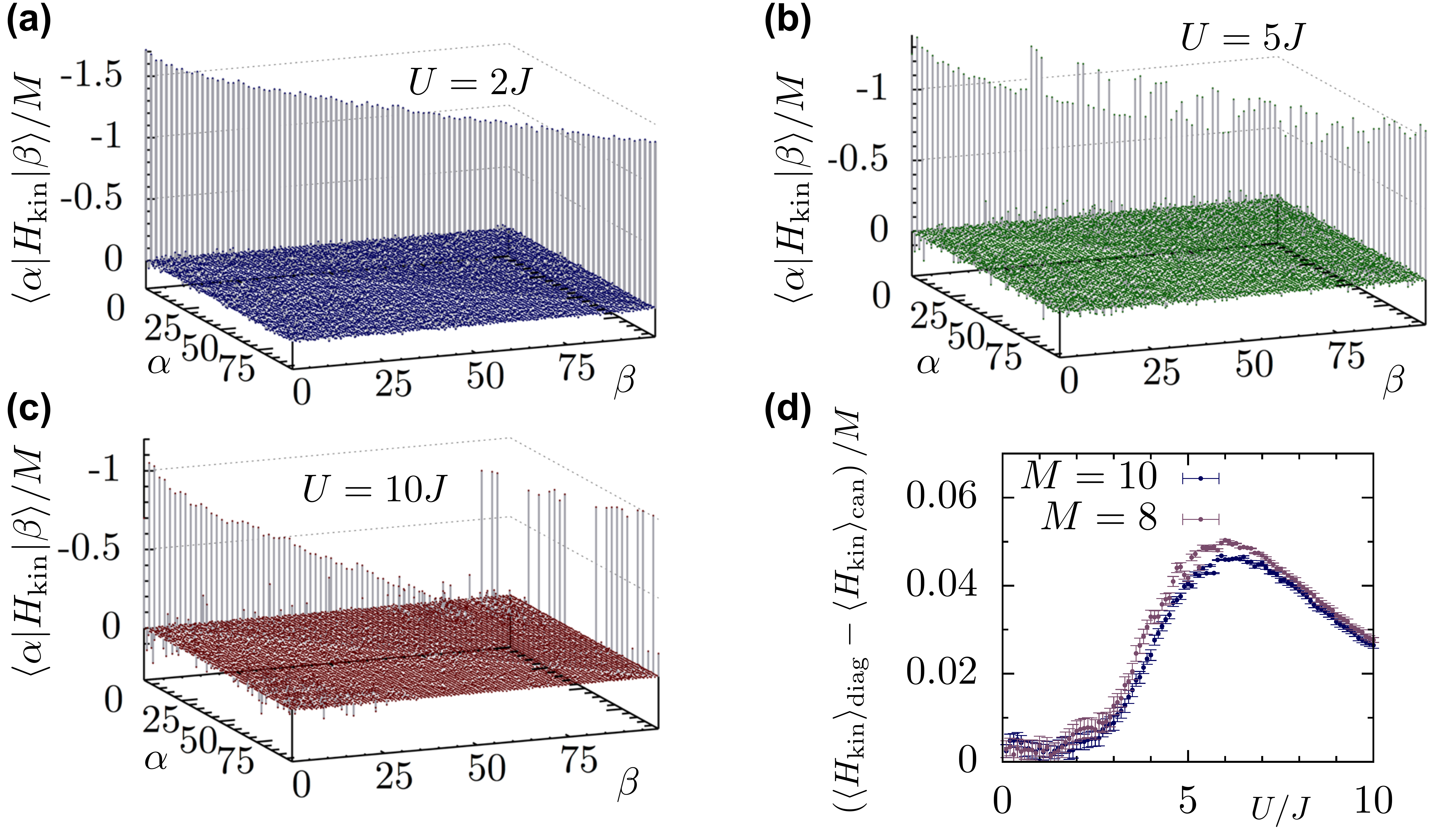}
\caption{{\em Eigenstate thermalization of spontaneous emissions}. (a)-(c) The matrix-elements of the kinetic energy ($N=M=10$),  $\bra{\alpha} H_{kin} \ket{\beta}/M$, in the energy eigenbasis (sorted in ascending order according to their energy). Different panels are for increasing on-site interaction $U$. Only in the SF regime ($U \lesssim 3.25(5) J$ \cite{Schonmeier-Kromer2014}) the diagonal elements vary smoothly with $\alpha$, and allow for eigenstate thermalization. (d) The difference of the expectations between the diagonal and the canonical ensemble corresponding to the same mean energy after a single spontaneous emission at site $i=M/2$ in a system of $M=8$ and $M=10$ sites (errorbars are due to the interpolation to find the canonical ensemble corresponding to the correct energy). Although finite size effects are large the change in behavior at the quantum phase transition is evident.}
\label{fig:eth_example} 
\end{figure}

In our situation, spontaneous emission only adds a small amount of energy to the system. Thus, the question whether this energy thermalizes or not will be determined  by the {\em low-energy} spectrum. Thus, we can use an exact numerical diagonalization to obtain the lowest energy eigenstates in systems with $N=M=10$ sites and particles. As an example, in panels (a)-(c) of Fig.~\ref{fig:eth_example} we show the matrix elements $\langle \alpha| \hat O | \beta \rangle$ for the kinetic energy in the system. As observable we use the kinetic energy $\hat O= H_{\rm kin}$. The eigenvalues are ordered ascending according to their energy.  We find that only in the SF case of $U=2J$, the diagonal elements vary smootly as function, while the off-diagonal elements are zero. This verifies the findings of Ref.~\cite{schachenmayer_spontaneous_2014}, and is consistent with the breakdown of the low-energy thermalization when entering the MI regime.

In addition we plot a direct comparison of the two expectation values in Fig.~\ref{fig:eth_example}d. We can calculate the expectation value on the one hand from the diagonal ensemble and on the other hand from a canonical ensemble with the same mean-energy. As initial state we choose the state after a single spontaneous emission on site $i=M/2$. Although finite size effects play an important role in this small system, it is already obvious that a discrepancy between the two ensembles arises when entering the MI regime. As was found by t-DMRG calculations in \cite{schachenmayer_spontaneous_2014}, this also holds for larger systems. In the next section we will  calculate the low-lying energy eigenstates exactly analytically in the limit of large interactions in the thermodynamic limit.

\section{Dynamics in the strongly interacting limit}
\label{sec:dh}

In this section we will discuss how we can understand the dynamics analytically for the case of strong interactions. In the extreme case of infinite $U$, at unit filling the particles will be in an idealized Mott insulator state, i.e., $\ket{\psi_{U\to \infty}}=\prod_n |1\rangle_n$. This state is an eigenstate of the jump operator with eigenvalue one, so that light scattering leaving particles in the lowest band does not change the state or introduce any extra energy to the system. 

\subsection{Hard-core bosons}

Going beyond this trivial limit, one can use an approximation in which we assume that the bosons are hard-core bosons (HCBs) \cite{rigol_universal_2004}. The idea is to restrict the Hilbert space to local states with a maximum of one particle per site. Mathematically, this can be achieved by using the standard Bose-Hubbard model~(\ref{eq:bh}) while additionally imposing fermionic anti-commutation relations to the creation/annihilation operators for bosons on the same site, $\{b_i,b_i^\dag\}=1$. These particles are not real fermions, since they still commute on different sites, $[b_i,b_j^\dag]=1$. However this can be fixed by a Jordan Wigner transformation \cite{jordan_1928}
\begin{equation}
	b_i = \prod_{\alpha =1}^{i-1} e^{{\rm i} \pi c_\alpha^\dag  c_\alpha} c_i,
\end{equation}
where the new quasi-particles now are real fermions, $\{c_i,c_j^\dag\}=\delta_{i,j}$. In this picture the Hamiltonian becomes a Hamiltonian for non-interacting fermions:
\begin{equation}
H=-J\sum_{i} (c_i c_{i+1}^\dag + c_i^\dag c_{i+1}),
\end{equation}
and the ground-state takes the product form $\ket{\psi_G} = \prod_{n}^N \sum_{i}^N P_{i,n} \ket{i}$ (where $\ket{i}$ denotes the state of a particle at site $i$). The jump-operators -- as is true for all site-local operators  -- are unaffected by the Jordan-Wigner transformation and so in this transformed notation, $b_i^\dag b_i \rightarrow c_i^\dag c_i$. 

If the system has fewer particles than lattice sites, spontaneous emissions will give rise to heating in the lowest band, as the states are not eigenstates of $c_i^\dag c_i$, which does not commute with the Hamiltonian. Simple results in this case are discussed in Ref.~\cite{Daley2014}.  However, in the special case of unit filling, which we would like to treat here, the only state described by this simple HCB form is again the trivial state with a single particle on each site, which is an eigenstate of all operators $c_i^\dag c_{i}$ and $c_i^\dag c_{i+1}$. It is then clear that we need to go beyond this treatment in order to understand heating in a Mott Insulator state with finite $U$.

\subsection{Doublon/Holon calculation}

To study thermalization of spontaneous emission we have to go to the next order in the approximation and allow for states with $0$, $1$, and $2$ particles per site. It turns out that also in this case approximate analytical calculations can be derived \cite{barmettler_propagation_2012}.  Assuming that the state of the system remains close to the Mott insulator state with $\bar n=1$, to first order excitations will be given by doubly occupied sites on the one hand, and holes on the other. It is now possible to introduce creation/annihilation operators with a vacuum given by the ideal MI state $\ket{\rm vac}_i=\ket{1}_i$ as 
\begin{eqnarray}
	\tilde d^\dag_i \ket{\rm vac}_i &=\ket{2}_i \\
	\tilde h^\dag_i \ket{\rm vac}_i &=\ket{0}_i
\end{eqnarray}
Since there can only be one hole or doublon on each site these quasi-particles must obey the hard-core constraint, which can again be expressed with on-site anti-commutation relations $\{\tilde d_i,\tilde d_i^\dag\}=1=\{\tilde h_i, \tilde h_i^\dag\}$. Note that in principle one also has to add a constraint in order not to have a doublon and holon on the same site. However, based on the assumption that the total number of doubly-occupied sites and holes remains low we will neglect these terms, as was originally done in Ref.~\cite{barmettler_propagation_2012}. As in the case of the HCB model one can now turn the quasi-particles into proper fermions via the Jordan-Wigner transformation
\begin{eqnarray}
	\tilde d_i &= \prod_{\alpha =1}^{i-1} e^{{\rm i} \pi d_\alpha^\dag  d_\alpha} d_i \\
	\tilde h_i &= \prod_{\alpha =1}^{i-1} e^{{\rm i} \pi h_\alpha^\dag  h_\alpha} h_i.
\end{eqnarray}
In the picture of these fermionic quasi-particles, the Hamiltonian reads
\begin{eqnarray}
	H_{dh}&= U \sum_i d_i^\dag d_i - J \left (2\sum_i d_i d^\dag_{i+1} + \sum_i  h_i h^\dag_{i+1} + {\rm h.c.} \right) \nonumber \\
	&- \sqrt{2} J  \left( \sum_i d^\dag_i h^\dag_{i+1} - \sum_i h_i d_{i+1}   \right).
\end{eqnarray}
A discrete Fourier transformation into quasi-momentum space ($x_j=$ $\frac{1}{\sqrt{M}}$ $\sum_q$ $e^{{\rm i}jqa} x_q$, for the operator $x_j$, $q \in \{0,2\pi/M,2\times 2\pi/M\dots 2\pi \}$) gives the quadratic Hamiltonian (up to a constant)
\begin{align}
\tilde H_{dh} = \sum_q
\begin{pmatrix}
d_q^\dag &
h_{-q}
\end{pmatrix}
\begin{pmatrix}
 E_d(q) & i E_{dh}(q) \\
-i E_{dh}(q) & -E_h(q)
\end{pmatrix}
\begin{pmatrix}
d_q\\ 
 h_{-q}^\dag
\end{pmatrix}.
\end{align}
The corresponding dispersion relations are given by $E_d(q)=-4J\cos(q) + U$, $E_h(q)=-2J\cos(q)$, and $E_{dh}(q)=2J\sqrt{2} \sin(q)$. This Hamiltonian can be diagonalized straightforwardly by a Bogoliubov transformation
\begin{widetext}
\begin{align}
H &= \sum_q
\begin{pmatrix}
d_q^\dag &
h_{-q}
\end{pmatrix}
U_B(q) U_B(q)^\dag
\begin{pmatrix}
 E_d(q) & i E_{dh}(q) \\
-i E_{dh}(q) & -E_h(q)
\end{pmatrix}
U_B(q) U_B(q)^\dag
\begin{pmatrix}
d_q\\ 
 h_{-q}^\dag
\end{pmatrix}\\
&= \sum_q
\begin{pmatrix}
d_q^\dag &
h_{-q}
\end{pmatrix}
U_B(q)
\begin{pmatrix}
 \epsilon_d(q) & 0 \\
0 & -\epsilon_h(q)
\end{pmatrix}
 U_B(q)^\dag
\begin{pmatrix}
d_q\\ 
 h_{-q}^\dag
\end{pmatrix} \\
&\equiv \sum_q
\begin{pmatrix}
c_{d,q}^\dag &
c_{h,-q}
\end{pmatrix}
\begin{pmatrix}
 \epsilon_d(q)& 0 \\
0 & -\epsilon_h(q)
\end{pmatrix}
\begin{pmatrix}
c_{d,q}\\ 
 c_{h.-q}^\dag
\end{pmatrix} \\
&=
 \sum_q
\epsilon_d(q) c_{d,q}^\dag c_{d,q}
 +
 \epsilon_h(q) c_{h,-q}^\dag c_{h,-q}
\end{align}
\end{widetext}
where the new fermionic quasi-particles $c_{d/h}$ are linear combinations of the $d_q$ and $h_q$, i.e.~are correlated pairs of doublons and holons. The dispersion relations of the particles are given by (assuming  $U>6J$)
\begin{align}
\epsilon_{d/h}(q)= \pm J \cos(q) + \frac{1}{2} \sqrt{32
  J^2 \sin ^2(q)+[U-6 J \cos (q)]^2}.
  \label{eq:dh_disp}
\end{align}
The ground state in this approximation is the vacuum $| {\rm vac} \rangle $, which is
defined by  $c_{d/q}\ket{\rm vac}=0$ for all $q$.

\subsection{Spontaneous emission in the doublon/holon picture}

\begin{figure}[t]
\begin{center}
\includegraphics[width=1\columnwidth]{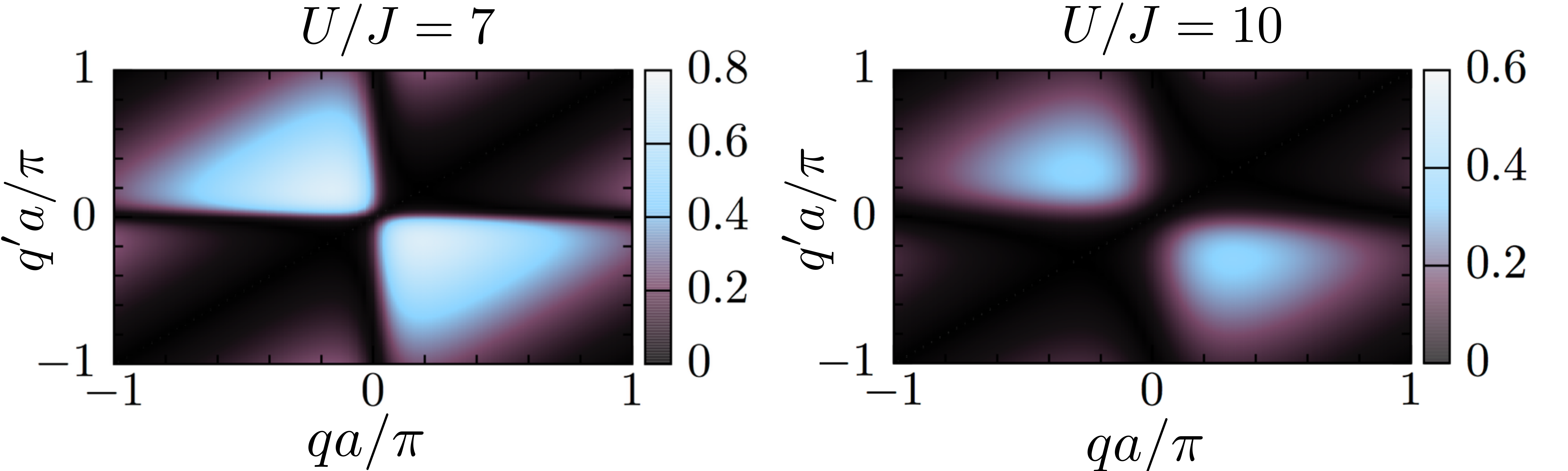}
\caption{The quasimomentum distribution of the initial state that is created by a spontaneous
  emission in the doublon-holon model for $U=7J$ (left panel) and
  $U=10J$ (right panel).\label{fig:dh_wp}}
\end{center}
\end{figure}

Using the inverse transformation, $(c_{d,q}^\dag , c_{h,-q})\,
U_B^\dag(q) = (d_{q}^\dag , h_{-q}) $, we can express the the local
particle number operator for site $m$, $n_m$ in the Bogoliubov
frame. Applying this operator to the ground-state, we find
\begin{align}
&n_m |{\rm GS} \rangle =\nonumber | {\rm vac} \rangle \\& + \frac{1}{M} \sum_{q,q'} (u_{q'} v_q - u_q v_{q'}) c^\dag_{d, q'}c^\dag_{h, -q} e^{ima(q-q')}| {\rm vac} \rangle,
\end{align}
where we introduced the matrix elements of the transformation as $u_q
= U_B(q)[1,1]$, and $v_q = U_B(q)[2,1]$. The time-evolved state after the
jump is therefore

\begin{align}
\label{eq:jump_state}
\sqrt{\mathcal{N} }|\psi (t) \rangle &= | {\rm vac} \rangle \nonumber \\ 
&+ \frac{1}{M} \sum_{q,q'}e^{ima(q-q')} (u_{q'} v_q - u_q v_{q'}) \nonumber \\ 
&\times e^{-it[\epsilon_d(q') - \epsilon_h(q)]} c^\dag_{d, q'}c^\dag_{h, -q} | {\rm vac} \rangle
,
\end{align}
where $\mathcal{N}\equiv \langle {\rm GS} | n_m^2 | {\rm GS} \rangle
$. The initial probablity-distribution of the wavepacket $w(q,q')\equiv u_{q'} v_q - u_q v_{q'}$  is
proportional to
\begin{align}
 |w(q,q)'|^2\propto \sin\Bigg[{\arctan\left(\frac{4\, \sqrt{2}\, J\, \sin\left({q'}\right)}{U - 6\, J\, \cos\left({q'}\right)}\right)} \nonumber \\- {\arctan\left(\frac{4\, \sqrt{2}\, J\, \sin\left({q}\right)}{U - 6\, J\, \cos\left({q}\right)}\right)}\Bigg]^2
\end{align}
and examples for $U=7J$ and $U=10J$ are depicted in
Fig.~\ref{fig:dh_wp}. The wavepackets are peaked at opposite momenta and
therefore, the jump creates holes and double occupations which move in
the opposite direction. For example, for $U/J=10$, the wavepackets are
peaked at $qa=-q'a\approx 0.30\pi$ and we can calculate the
group-velocity from \eqref{eq:dh_disp} as $\frac{d}{dqa}
\epsilon_{d}(q)|_{0.3\pi} \approx 3.75 a/J$.

We can test this analytical result with an exact t-DMRG time-evolution calculation after a single jump. We compare the dynamics after a jump in the center of a system with $N=M=48$ site in Fig.~\ref{fig:dh_wp_dyn}a. There we plot the time-evolution of the number of doubly occupied sites as a function of time. The white dashed line in Fig.~\ref{fig:dh_wp_dyn}a indicates the light-cone according to the analytically expected group-velocity and is in excellent agreement with the numerical calculations.


\subsection{Behaviour of the kinetic energy and similar variables}

\begin{figure}[t]
\begin{center}
\includegraphics[width=1\columnwidth]{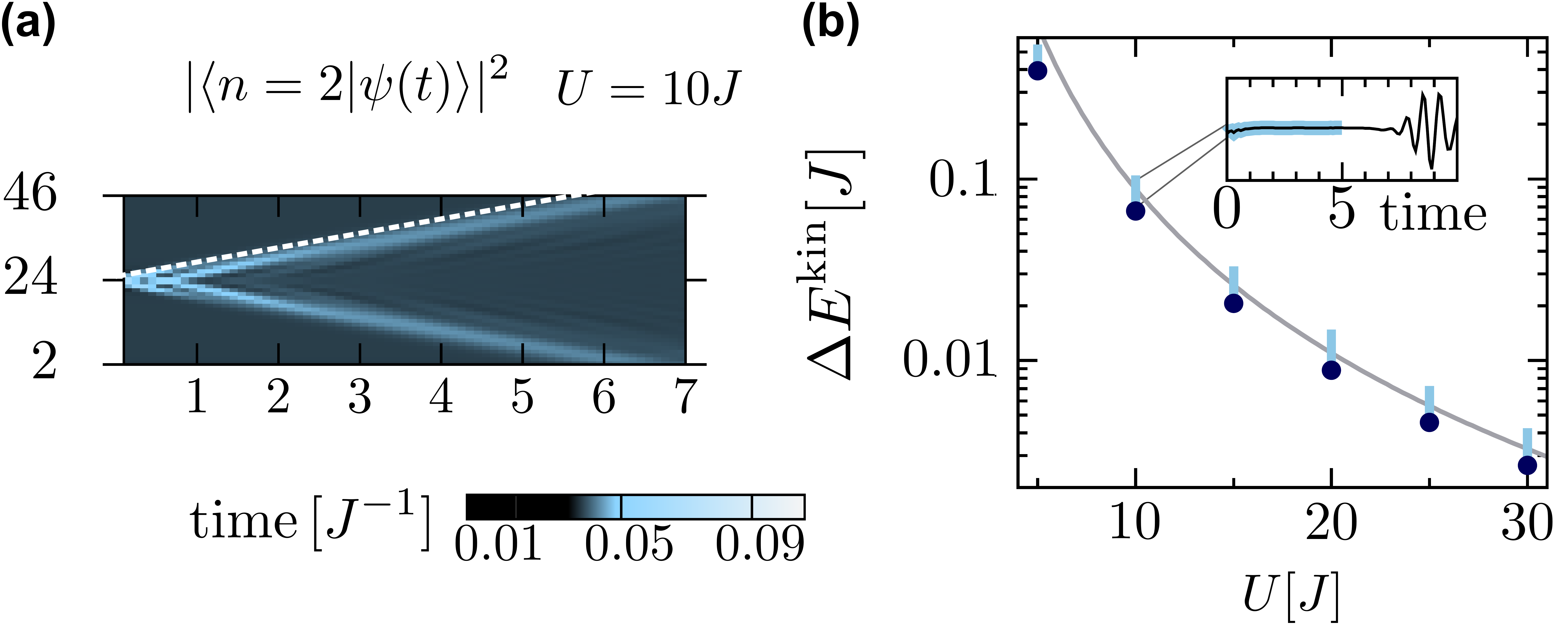}
\caption{{\em Dynamics after a single spontaneous emission} (a) Time-evolution of the probability of doubly occupied sites after a single spontaneous emission in the center of a
  $48$ site system. This observable reveals 
  doublon-holon pair wave-packets, which propagate freely through the system in opposite directions until they reach the boundary
  (dashed white line: analytical result for the light-cone in the thermodynamic limit). (b) Comparison of
  the corresponding time-evolution of the kinetic energy energy difference to the
  ground-state. The grey line shows the time-independent analytical result for the thermodynamic limit. The dots show the numerical result right after the jump and the blue line indicates the values that develop before boundary effects become important ($tJ<5$, see inset).  (t-DMRG calculation converged with $D=256,512$, $n_m=6$) 
  \label{fig:dh_wp_dyn}}
\end{center}
\end{figure}

We now consider a specific type of observable, such as the kinetic energy, which in quasi-momentum space and in the Bogoliubov frame can be written as 
$O = \sum_q  O_q$ with
\begin{align}
O_q&\equiv O_{11}(q) c_{d,q}^\dag c_{d,q} + O_{12}(q) c_{d,q}^\dag c_{h,-q}^\dag \nonumber\\
&+O_{21}(q) c_{h,-q} c_{d,q} + O_{22}(q) c_{h,-q} c_{h,-q}^\dag
,
\label{eq:qobs}
\end{align}
with matrix elements $O_{ij}$ for each $q$. Calculating the
expectation value of this observable with the time-evolved state gives
\begin{align}
\label{eq:op_expc_res}
& \langle \hat O \rangle(t) = 
\frac{1}{\mathcal{N}} \sum_q O_{22} \nonumber \\
&+ \frac{1}{\mathcal{N}} \frac{1}{M^2} \sum_{q,q'} \left( O_{11}(q') - O_{22}(q) + \sum_p O_{22}(p) \right) |w(q,q')|^2
\end{align}
The important points to note are that all cross-terms corresponding to the $O_{12}$ and $O_{21}$ part of the operator disappear. This is because terms such as
$ \langle {\rm vac} |c_{h,-p} c_{d,p}  c_{d,q'}^\dag c_{h,-q}^\dag | {\rm vac} \rangle  = \delta_{p,q'} \delta_{p,q}$ become zero due to the symmetric form of the specific wave-packet that is created in the jump, $w(q,q')\equiv u_{q'} v_q - u_q v_{q'}$, $w(p,p)=0$. Similarly, Kronecker deltas in the terms of the form $ \langle {\rm vac}| c_{h,-r} c_{d,r'} c_{d,p}^\dag c_{d,p} c_{d,q'}^\dag c_{h,-q}^\dag | {\rm vac} \rangle 
=   \delta_{p,q'} \delta_{r',p} \delta_{r,q}$ lead to a cancellation of time-dependent terms. Thus, we find that any observable of this type is time-independent after the spontaneous
emission. This is a remarkable result, given the fact
that the state \eqref{eq:jump_state} is not an eigenstate of the
Hamiltonian. The effect of a jump operator here is that it puts the
system  into a state whose density matrix is diagonal
in the space of single doublon-holon pair excitations, for which case these specific observables don't evolve in time. In particular, the expectation value of observables of the form~\eqref{eq:qobs} can be effectively expressed by a density matrix -- $\langle  O \rangle = {\rm tr} (\rho O)$ -- of the form
\begin{align}
&\rho = {\lambda_{\rm vac} | {\rm vac} \rangle\langle {\rm
    vac}| +\sum_{q,q'} \lambda_{q,q'} | q,q' \rangle\langle q,q'|},
\end{align}
where  $| q,q' \rangle \equiv c_{d,q'}^\dag c_{h,q}^\dag | {\rm vac}\rangle$. Thus, we conclude that after a single spontaneous emission, observables of the form \eqref{eq:qobs} can immediately be described by the diagonal ensemble, i.e.~the effective density matrix for those observables takes a diagonal form in the basis low-lying energy states.

As an example, let us now consider the kinetic energy operator, which can be written as
\begin{align}
 O_q= T_q=
\begin{pmatrix}
-4 J \cos[q] & -2i\sqrt{2} J \sin[q] \\ 2i\sqrt{2} J \sin[q]  & 2 J \cos[q]
\end{pmatrix}
.
\end{align}
In the thermodynamic limit, we can replace the sum over quasi-momenta with the integral,
$(1/M) \sum_q \rightarrow \int d(qa)/(2\pi)$, which yields
\begin{align}
\langle \hat H_{\rm kin } \rangle (t) = - 8 M \frac{J}{U} - (48 M+88) \left(\frac{J}{U}\right)^3 + \mathcal{O}\left[ \left(\frac{J}{U}\right)^5
 \right].
\end{align}
We compare this analytical result to our time-dependent t-DMRG calculations This is in excellent agreement with our numerical results even for
relatively small $U/J$ as shown in Fig.~\ref{fig:dh_wp_dyn}b.

\subsection{Eigenstate expectation values}

\begin{figure}
 \includegraphics[width=1\columnwidth]{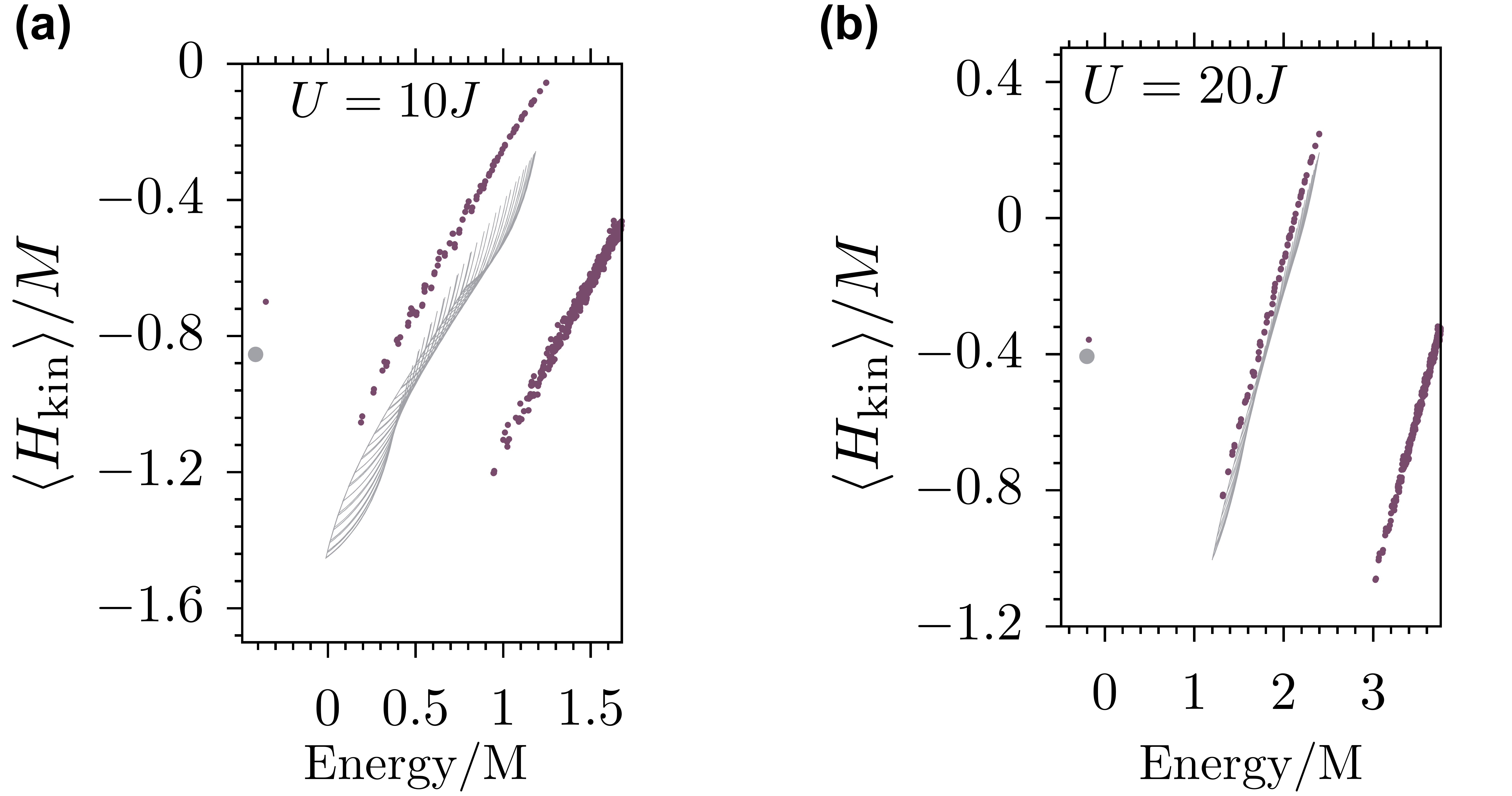}
\caption{{\it Eigenstate expectation values of the kinetic energy --} Violet dots are results from an exact diagonalization calculation in a system with $N=M=10$ particles/sites. Panel (a) is for $U=10J$, panel (b) for $U=20J$. The grey dot is the result for the ground-state kinetic energy from the analytical doublon-holon calculation in the thermodynamic limit. Grey lines are expectation values for the first branch of single doublon-holon pair excitations. For increasing $U$ the numerical results approach  the analytical result as expected (differences arise from contributions with more than two-particles per site and finite size effects).}
\label{fig:eth_comp} 
\end{figure}

We can check the eigenstate expectation values of the kinetic energy analytically. In Ref.~\cite{schachenmayer_spontaneous_2014}, it was found that, once one enters the MI regime, these do not coincide with the expectation values for Boltzmann distributions with the corresponding energies anymore (cf.~Section~\ref{sec:eth}). Now we can check whether these agree with our analytical calculation. Therefore, using exact diagonalization we compute the $500$ lowest eigenstates in a system of $N=M=10$ sites. These are shown as dots in Fig.~\ref{fig:eth_comp}a,b for the case of $U=10J$ and $U=20J$, respectively. In addition we add the values we obtain analytically for the ground state (grey dot) and the branch of single doublon-holon pair excitations (grey lines).  We find good agreement for large interactions. Since the first branch converges towards the analytical calculation, we conclude that this branch indeed due to the single doublon-holon pair excitations, which is further verified by the fact that it contains $M^2-M$ points. The second branch must thus correspond to two pair excitations. Already soon after entering the MI $U\sim 5$ the gap and branches start to appear \cite{schachenmayer_spontaneous_2014}. This indicates that the doublon-holon calculations are indeed a good approximation even for regimes that are not particularly deep in the MI phase.

\section{Summary and Outlook}
\label{sec:summary}

In summary, we have investigated the non-equilibrium dynamics of bosons in an optical lattice in the presence of spontaneous emission events. We show clearly parameter regimes in which the system relaxes to thermal distributions for simple quanitites (especially the quasimomentum distribution), and others where it relaxes on short timescales to non-thermal values. We can understand this behavior by applying eigenstate thermalization considerations to small systems for which we can perform exact diagonalization calculations. We also show that we can understand the dynamics in the strongly interacting limit well in terms of propagating doublon-hole pairs, which are analytically tractable. 

All of these results are expected to be directly observable in ongoing experiments. In addition to the possibility to measure quasimomentum distributions, the propagation of doublon-holon pairs as investigated here is a key signature for the effects of spontaneous emissions that could be measured in quantum gas microscope experiments with single-site resolution in ways that were previously implemented for quenches within the MI regime \cite{cheneau_light-cone-like_2012}. Measurements of this type could be used as a diagnostic tool for heating of many-body states in optical lattices, which could in turn be used to improve the robustness of quantum simulators. In the future, these studies can be continued towards thermalization for fermions in optical lattices undergoing spontaneous emissions\cite{Bernier2013,Sarkar2014,Bernier2014}, and the incorporation of partial thermalization from excitations to higher Bloch bands.

\begin{acknowledgements}
We would like to thank I.~Bloch, W.~Ketterle, S.~Langer, H.~Pichler,
U.~Schneider, and P.~Zoller for discussions. This work was supported in part by AFOSR grant
FA9550-13-1-0093, by a grant from the US Army Research Office with
funding from the DARPA OLE program, and by the Aspen Center for Physics support under NSF grant 1066293. Computational resources were
provided by the Center for Simulation and Modeling at the University
of Pittsburgh. 
\end{acknowledgements}

\bibliographystyle{spphys}       

\bibliography{therm}

\end{document}